\documentstyle[aps,epsf]{revtex}
\title{Quasi-Solitons in Dissipative Systems and Exactly Solvable Lattice Models}
\author{Yuji Igarashi,~~Katsumi Itoh}
\address{Faculty of Education, Niigata University, Niigata 950-2181, Japan}
\author{Ken Nakanishi}
\address{Department of Physics, Nagoya University, Nagoya 464-0814, Japan}
\author{Kazuhiro Ogura}
\address{Hosei University Daini High Schools, Kawasaki 211-0031, Japan}
\author{Ken Yokokawa}
\address{Faculty of Education, Niigata University, Niigata 950-2181, Japan}

\date{\today}

\begin{document}
\twocolumn[\hsize\textwidth\columnwidth\hsize\csname
@twocolumnfalse\endcsname
\maketitle
\begin{abstract}

A system of first-order differential-difference equations with time lag
describes the formation of density waves, called as quasi-solitons for
dissipative systems in this paper. For co-moving density waves, the
system reduces to some exactly solvable lattice models. We construct a
shock-wave solution as well as one-quasi-soliton solution, and argue
that there are pseudo-conserved quantities which characterize the
formation of the co-moving waves. The simplest non-trivial one is given
to discuss the presence of a cascade phenomena in relaxation process
toward the pattern formation.

\end{abstract}
\hspace*{1.8cm}
PACS numbers:
05.45.Yv, 05.45.-a, 04.20.Jb, 02.60.Cb

\vskip1.9pc]
\narrowtext  



Recently three of the present authors have found exact
solutions\cite{Igarashi} of the first-order differential-difference
equations\cite{hasebebando}
\begin{eqnarray} 
{\dot x}_{n}(t+\tau)&=&V[\Delta x_{n}(t)]\nonumber\\
&\equiv& \xi+\eta
\tanh\left[\left(\frac{\Delta x_{n}(t)-\rho}{2A}\right)\right],
\label{DDEOM} 
\end{eqnarray} 
where $\tau$ is a time lag and $\Delta x_{n}(t)= x_{n-1}(t) - x_{n}(t)$
the forward difference on discrete indices. This system exhibits
temporal-spatial density patterns of constituent elements, and can be
used as a one-dimensional model for a variety of the co-operative
phenomena such as congested flows of traffic, flows of granular
particles, and possibly some chemical as well as ecological
problems. The periodic solutions\footnote{An exact solution has been
found also by Hasebe, Nakayama and Sugiyama\cite{hasebe}.  It is a
special case of our solutions, obtained by a Landen transformation of
elliptic functions.} we have obtained precisely describe the steady
profile density waves observed in numerical simulations. The density
waves are locally stable and behave as soliton-like excitations. They
are however neither necessarily ``solitary'' nor independent each other. 
It will be suitable to distinguish those appeared in dissipative systems
from the solitons associated with integrable systems.  We call the new
objects quasi-solitons (QS's).

It is remarkable that the system (\ref{DDEOM}) may exhibit a kind of
partial integrability upon the spatial-temporal pattern formation.  The
idea has been suggested by three of us\cite{Igarashi} and also by the
authors of ref.\cite{hasebe}, who have written eqs.(\ref{DDEOM}) in a
form which has a lattice analog, the system of Hirota's self-dual LC circuit equations.

The purpose of this letter is to establish the connection between the
system (\ref{DDEOM}) and relevant lattice models, constructing some exact
solutions for an infinite ($-\infty< n < \infty$) system. Based on the
reduction to the Hirota lattice\cite{Hirota}, we construct a shock-wave
solution and one-QS solution. To get an insight into QS-solutions, we
also transform eqs.(\ref{DDEOM}) into the
Hirota-Satsuma lattice model\cite{HirotaSatsuma}, an extended version of
the Hirota's model. One of the most important implications of the
lattice model correspondence is that one may construct pseudo-conserved
quantities in the dissipative system which quantitatively characterize
relaxation processes toward the pattern formation.

The lattice model correspondence applies only to the density waves which
have the following properties:\\
\noindent
(i) a common phase variable $ w(t) = \nu t - k n $;\\
\noindent
(ii) a linear dispersion relation\cite{Whitham},
\begin{eqnarray}
\nu  = k/(2 \tau).
\label{Whitham}
\end{eqnarray}
The system (\ref{DDEOM}) describing propagation of the co-moving waves
with these properties can be reduced to exactly solvable lattice
models. Actually, the connection with the Hirota lattice is readily seen
by defining
\begin{eqnarray}
M_{n}(t)= \tanh\left[(\Delta x_{n}(t)-\rho)/2A\right], 
\label{defM}
\end{eqnarray}
and using (\ref{Whitham}) to replace the time shift $t \to t-\tau$ by the
subscript shift $n \to n+1/2$: $w(t-\tau)=\nu t - k (n+1/2)$.  Making
the total degrees of freedom twice of those of the original system
(\ref{DDEOM}), one obtains the Hirota lattice equations as reduced equations
\begin{eqnarray}
2\tau_{c}~{\dot M}_{n} =\left(1 - M_{n}^2\right)\left(M_{n-\frac{1}{2}}-M_{n+\frac{1}{2}}\right).
\label{Hirota}
\end{eqnarray}
where $\tau_{c}= A/\eta$ is a constant.  As for the basic variables, we
may take an ansatz
\begin{eqnarray}
x_{n}(t) = A~{\ln}~\varphi_{n}(t) ~+~ Ct~-n h, 
\label{ansatz}
\end{eqnarray}
where $h$ and $C=V(h)$ are constants.  $M$ is expressed by $\varphi$'s as
\begin{eqnarray}
M_{n}&=& \frac{\zeta \varphi_{n-1} - \varphi_{n}}{\zeta \varphi_{n-1} +
\varphi_{n}}~,~~~~ \zeta= \exp [(h-\rho)/A].  
\label{Mphi}
\end{eqnarray}
We give some solutions using these variables.

\noindent
(1)~~{\it Shock-wave solution} is given by
\begin{eqnarray}
\varphi_{n}&=& 1~+~\exp (-2w + k), \nonumber\\
M_{n}&=& \tanh(k/2)~\tanh w,
\label{shock1}
\end{eqnarray}
where the parameter should satisfy
\begin{eqnarray} 
h = \rho + A~k,~~~~ \frac{\tau_c}{2\tau}~k= \tanh (k/2).
\label{shock2}
\end{eqnarray}
The solution (\ref{shock1}) yields a kink in the velocity variable
\begin{eqnarray}
{\dot x}_{n}= A\nu\left[\tanh(w-k/2)~-1\right]~+~C,   
\label{kink}
\end{eqnarray}
which gives an interpolation between a uniform flow with the velocity
$C-2A\nu$ at $n=-\infty$ and that with $C$ at $n=\infty$.  Since the
transition between two flows takes infinite time, the solution
(\ref{shock1}) describes an asymptotic trajectory\cite{nakanishi} in phase
space.

\noindent
(2)~~{\it One-QS solution} takes of the form
\begin{eqnarray}
\varphi_{n}&=& \frac{1~+~\exp (-2w + k -2\delta)}{1~+~\exp (-2w + k
+2\delta)},\nonumber\\
M_{n}&=& \frac{\zeta -1}{\zeta
+1}~+~\frac{2\zeta}{(\zeta+1)^2}~u_{n}^{QS},\nonumber\\
u_{n}^{QS}&=& \frac{\sinh(2\delta)\sinh k}{\cosh^2 w + \sinh^2 \delta}, 
\label{one-QS}
\end{eqnarray}
where 
\begin{eqnarray}
\frac{k}{\sinh k}&=& \frac{\tau}{\tau_c}\frac{4\zeta}{(\zeta+1)^2},
\label{one-QS1}\\
 \tanh (k/2) &=& - \frac{\zeta -1}{\zeta +1}~\tanh(2\delta).
\label{one-QS2}
\end{eqnarray}
This solution describes a pair of kink and anti-kink 
\begin{eqnarray}
{\dot x}_{n}&=& A\nu\left[\tanh(w-k/2+\delta)-\tanh(w-k/2-\delta)\right]\nonumber\\
&&+C.
\label{kinkpair}
\end{eqnarray}

We have obtained above a shock wave solution as well as one-QS solution. 
To get a further insight into QS solutions, we rewrite the reduced equations as
those of a model of Hirota and Satsuma, who constructed exact solutions
from those of the Toda lattice using the B{\"a}cklund transformation
technique. We show that this formalism gives the same one-QS solution as
that of (\ref{one-QS}).  To this end, we define new variables by
\begin{eqnarray}
\frac{2\zeta}{(\zeta +1)^2}~u_{n}= M_{n} - \frac{\zeta - 1}{\zeta + 1},
\label{udef}
\end{eqnarray}
and a rescaled time variable
\begin{eqnarray}
\frac{2\zeta}{\tau_c (\zeta +1)^2}~t = s.
\label{tscale}
\end{eqnarray}
These obey then the reduced equations 
\begin{eqnarray}
\frac{d u_{n}}{d s}=\left[1 - \frac{\zeta -1}{\zeta +1}~u_{n} -
\frac{\zeta~u_{n}^2}
{(\zeta +1)^2}\right]\left(u_{n-\frac{1}{2}}- u_{n+\frac{1}{2}}\right). 
\label{ueq}
\end{eqnarray}
The system (\ref{ueq}) is exactly the same as the $(u,~v)$ system of the
Hirota-Satsuma lattice,
\begin{eqnarray}
\frac{d u_{n}}{d s} &=& \left[\alpha + \beta_{1} u_{n} + \frac{(\beta_{1}^2 -1)
u_{n}^2}{4 \alpha}\right](v_{n-\frac{1}{2}}- v_{n+\frac{1}{2}}), 
\nonumber\\
\frac{d v_{n}}{d s}&=& \left[\alpha^{-1} + \beta_{2} v_{n} + \frac{\alpha (\beta_{2}^2 -1)
v_{n}^2}{4}\right](u_{n-\frac{1}{2}}- u_{n+\frac{1}{2}}).
\nonumber
\end{eqnarray}
with $\alpha =1,~~\beta_{1}=\beta_{2}=-(\zeta -1)/(\zeta +1)$ and
$u_{n}=v_{n}$. The $v$ variables are redundant here. In the
Hirota-Satsuma construction, exact solutions are
given by
\begin{eqnarray}
u_{n}=\frac{d}{d s}\ln \left({\tilde f}_{n}/f_{n}\right),
\end{eqnarray}
where both $f_{n}$ and ${\tilde f}_{n}$ are solutions of the Toda lattice
equations:
\begin{eqnarray}
1~+~\frac{d^2}{d s^2} \ln f_{n}= \frac{f_{n+1}f_{n-1}}{f_{n}^2}.
\label{Toda}
\end{eqnarray}
The variables $u_{n}$ are suitable for solving the reduced equations
subject to the boundary conditions $u_{\infty} =u_{-\infty}=0$, which
describe a uniform flow with $\Delta x_{n} = h$ at spatial infinities.
As for one-soliton solution, we may take
\begin{eqnarray}
f_{n}&=& 1~+~\exp(2 W+ \phi), \nonumber\\
{\tilde f}_{n}&=& 1~+~\exp(2 W+ {\tilde \phi}), 
\label{os}
\end{eqnarray}
where  
\begin{eqnarray}
W &=&\Omega s - k n,  \nonumber\\
\Omega &=& \sinh k, \nonumber\\
\exp \phi &=& -\frac{\zeta -1}{\zeta +1}(1 + \cosh k) - \sinh k,
\label{DR} \\
\exp {\tilde \phi} &=&  -\frac{\zeta -1}{\zeta +1}(1 + \cosh k) + \sinh k.
\end{eqnarray}
This leads to
\begin{eqnarray}
u_{n}= \frac{\Omega \sinh[({\tilde \phi}-\phi)/2]}
{\cosh^{2}[W + ({\tilde \phi}+\phi)/4)] +
\sinh^{2}[({\tilde \phi}-\phi)/4]}, 
\label{usol}
\end{eqnarray}
which should be compared with $u_{n}^{QS}$ in (\ref{one-QS}). Consistency
in the dispersion relations requires $\Omega s = kt/(2\tau)$, which
yields the condition (\ref{one-QS1}). One obtains also $\delta = ({\tilde
\phi} - \phi)/4$, which leads to (\ref{one-QS2}). Therefore, after making
a constant shift of the time variable, two functions $u_{n}^{QS}$ and
$u_{n}$ become exactly the same.

It is well known that an exactly solvable lattice model has the same
number of conserved quantities as those of the degrees of freedom of the
model.  The lattice correspondence discussed above provides us with
pseudo-conserved quantities, which are generically time dependent, but
become conserved quantities upon the formation of spatial-temporal
patterns of density waves. We may construct them from conserved quantities
given by Wadati\cite{Wadati} for the Hirota lattice.  For a finite
system of (\ref{DDEOM}) subject to the periodic boundary condition
$x_{i}=x_{i+N}$, the simplest one is given by a quadratic form of the
$M$ variables:
\begin{eqnarray}
K(t)=
\sum_{n=1}^{N}\left[M_{n}(t)
\left(M_{n}(t+\tau)~+~M_{n+1}(t+\tau)\right)\right].
\label{conserv}
\end{eqnarray}
It is not a conserved quantity, but after the co-moving density waves are
generated, it reduces to
\begin{eqnarray}
{\cal K}(t)= \sum_{n=1}^{N}\left[M_{n}(t)
\left(M_{n-\frac{1}{2}}(t)~+~M_{n+\frac{1}{2}}(t)\right)\right].
\label{latticecons}
\end{eqnarray}
Using (\ref{Hirota}), one can see that the reduced ${\cal K}$ actually
conserves, $d{\cal K}/dt=0$.  Thus $K(t)$ characterizes relaxation
processes toward exact periodic solutions which act as ``attractors'' of
the system.  In Fig.1, we show the time evolution of $K(t)$ obtained
from a numerical simulation of (\ref{DDEOM}) for $N=20$.  The initial
state is an almost uniform flow with $K \sim 0.5$, the first plateau in
the figure.  The uniform flow with small perturbation becomes unstable, and there
appear density patterns with regions of high density where the constituent 
elements move slowly, and low-density regions where the velocities
of elements are high. The high density regions of congested flows are
viewed as clusters on the circuit. The final state with the maximum value for $K \sim
12.6$ is described by the exact periodic solution which contains one
cluster.  This is observed as the fourth plateau. In between, there
appear the second and third plateaus of $K \sim 7.7$ and $K \sim 10.1$ . 
These are the intermediate states well approximated by exact solutions
with three and two clusters.  Therefore, the relaxation process
starting from a perturbed uniform flow exhibits a cascade decay via
multi-cluster states. Our pseudo-conserved quantity $K(t)$
quantitatively characterizes this cascade phenomena.  There exist other
pseudo-conserved quantities which may take different values for
different pattern of flows, uniform or congested flows. Obviously we
may construct infinitely many kinds of such quantities with the same
nature for an infinite system $(-\infty <n < \infty)$.

\begin{figure}
    \leavevmode
    \epsfxsize=8cm
    \epsfbox{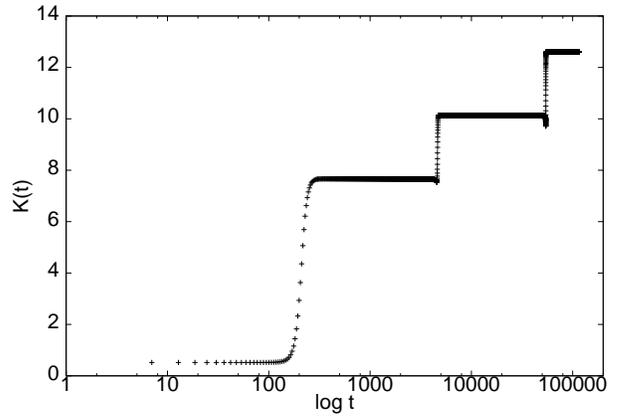} 
\caption{The time evolution of the
pseudo-conserved quantity $K(t)$ in a simulation.}
\end{figure}

We make several comments:\\
\noindent (a) Our dissipative system is shown to be related to
integrable systems and some solutions to the latter systems are
translated in the former as congested flows.  This connection, however,
does not imply the stability of the solutions in the original
dissipative system: they are related only after patterns are formed.  So
the stability of the solutions should be studied in the original system.
Actually we have confirmed that conditions in eq.(\ref{one-QS2}) allow us
to choose parameters to form typical stable patterns for congested flows
observed in simulations\cite{Igarashi}.  A shock-wave solution may be understood as the extreme
case of one-QS solution with $\delta \rightarrow \infty$, so the
stability of the latter implies the same nature of the former.
Therefore the solutions discussed in this paper are stable for suitable
range of parameters.\\  
\noindent (b) In simulations we have observed the initial condition dependence; 
a flow might develop to a congested flow or a uniform
flow depending on its initial condition.  If we could label initial
conditions with some of pseudo-conserved quantities, it would be a very
interesting application of the quantities. These applications will be discussed elsewhere.\\
\noindent (c) The dispersion relation (\ref{Whitham}) was first given by
Whitham\cite{Whitham} for Newell's model\cite{Newell} of traffic flows. It is a crucial condition
which makes a bridge between the dissipative system (\ref{DDEOM}) and
exactly solvable lattice models. We showed that this relation is a kind
of the integrability condition under which the system eq.(\ref{DDEOM})
admits periodic wave solutions\cite{Igarashi}.  Furthermore, Hasebe, Nakayama and
Sugiyama observed in numerical simulations that the relation in eq.(\ref{Whitham}) holds irrespective of details of the function $V(\Delta
x)$\cite{hasebe}. It is therefore a mysterious but non-trivial dynamical relation, whose derivation is challenging.\\
\noindent (d) It is difficult to find a two-QS solution of the system
(\ref{DDEOM}). The co-moving limit of the Toda lattice solutions in which
two-solitons propagate with a same velocity is not known.  Therefore, one may not construct a two-QS
solution from the Toda lattice solutions. It may not exclude, however,
the existence of such solution.  Actually, there exist periodic
solutions of eq.(\ref{DDEOM}) for finite as well as infinite systems
which do not solve the Toda equations. To find a two-QS solution, one
has directly to solve the reduced equations (\ref{Hirota}) or (\ref{ueq})
without referring to the Toda lattice solutions.  Analysis of the
co-moving limit in exact solutions obtained by Ablowitz and
Ladik\cite{Ablowitz} may give another insight into this problem. In this
paper, we have restricted ourselves to exact solutions of co-moving waves. We do
not know if multi-QS's with different velocities, observed in numerical
simulations, have corresponding analytic expressions. Discovery of
such solutions, if any, is also challenging.

We are grateful to  F.~Nakamura, K.~Tomisaka and T.~Otofuji for their assistance in computer simulation.


\end{document}